\documentclass[aps,10pt,pre,byrevtex,twocolumn,footinbib,showpacs,showkeys,longbibliography]{revtex4-1}
\usepackage{microtype}
\usepackage{graphicx}
\usepackage{amsmath}
\usepackage{amssymb}
\usepackage{color}
\definecolor{myblue}{rgb}{0.153,0.322,0.706}
\usepackage[colorlinks,linkcolor=myblue,urlcolor=myblue,citecolor=myblue,breaklinks=true]{hyperref}

\newcommand{\be}{\begin{equation}}
\newcommand{\ee}{\end{equation}}
\newcommand{\idf}{1\!\! 1}
\newcommand{\ra}{\rightarrow}

\newcommand{\reals}{\mathbb{R}}

\newcommand{\tA}{\tilde A}
\newcommand{\hj}{\hat\jmath}
\newcommand{\hrho}{\hat\rho}
\newcommand{\hu}{\hat u}
\newcommand{\hsig}{\hat\Sigma}
\newcommand{\diss}{\text{diss}}
\newcommand{\eq}{\text{eq}}

\begin{document}

\title{Process interpretation of current entropic bounds}

\author{Cesare Nardini}
\email{cesare.nardini@gmail.com}
\affiliation{Service de Physique de l'\'Etat Condens\'e, CNRS UMR 3680, CEA-Saclay, 91191 Gif-sur-Yvette, France}

\author{Hugo Touchette}
\email{htouchet@alum.mit.edu, htouchette@sun.ac.za}
\affiliation{National Institute for Theoretical Physics (NITheP), Stellenbosch 7600, South Africa}
\affiliation{\mbox{Institute of Theoretical Physics, Department of Physics, University of Stellenbosch, Stellenbosch 7600, South Africa}}

\date{\today}

\begin{abstract}
We show for Markov diffusion processes that the quadratic entropic bound, recently derived for the rate functions of nonequilibrium currents, can be seen as being produced by an effective process that creates current fluctuations in a sub-optimal way by modifying only the non-reversible part of the drift or force of the process considered while keeping its reversible part constant. This provides a clear interpretation of the bound in terms of a physical process, which explains, among other things, its relation to the fluctuation relation, linear response, and reversible limits. The existence of more general quadratic bounds, and related uncertainty relations, for physical quantities other than currents is also discussed.
\end{abstract}

\pacs{%
02.50.-r, 
05.10.Gg, 
05.40.-a 
}

\maketitle

\section{Introduction}


It has been found recently that large deviation rate functions characterizing the fluctuations of currents in nonequilibrium Markov processes are bounded above by a parabola determined only by the stationary current and stationary entropy production of the process considered \cite{pietzonka2015,gingrich2016,gingrich2017}. Such a general bound, referred to as the \emph{entropic bound}, is useful as it provides a simple Gaussian approximation of the full current distribution, which is in general very difficult to find analytically or numerically. Being an upper bound, it also implies a lower bound on the variance of the current, similar to the uncertainty relation in quantum mechanics, which shows that there is a physical trade-off between reducing fluctuations and dissipation \cite{pietzonka2015,gingrich2016,gingrich2017,barato2015b}. This has applications for studying the efficiency and accuracy of noisy reactions and processes performing biophysical tasks at the molecular level \cite{pietzonka2015,gingrich2016,gingrich2017,barato2015b,barato2015c,pietzonka2016,pietzonka2016b,barato2016,nguyen2016}.

In this note, we give a physical interpretation of the entropic bound for Markov diffusions by showing that it is associated with an effective diffusion for the fluctuations that modifies the non-reversible part of the drift of the original process considered. This interpretation follows from recent works on Markov processes conditioned on large deviations \cite{chetrite2013,chetrite2014,chetrite2015} and explains physically why the entropy bound is only an approximation of the true current rate function. It also provides, as we show, a basis for understanding how tight the bound is, why it satisfies the fluctuation relation symmetry, whether or not it is related to linear response \cite{gingrich2016}, and its applicability for reversible (equilibrium) processes. To conclude, we discuss the issue of deriving more general bounds and uncertainty relations for observables other than the current.

\section{Current large deviations}

We consider a general diffusion process $X_t$ in $\reals^n$ evolving according to the stochastic differential equation
\be
dX_t = F(X_t)dt+\sigma dW_t,
\label{eqsde1}
\ee
where $F$ is an $n$-dimensional vector field, called the drift, $W_t$ is an $m$-dimensional vector of independent Brownian motions, and $\sigma$ is an $n\times m$ matrix, taken to be constant for simplicity \footnote{The case where $\sigma$ is a function of the state can be treated similarly following the notations of \cite{chetrite2014}.}. We assume that the choice of $F$ and $\sigma$ is such that $X_t$ has a stationary probability density $\pi(x)$ solving the time-independent Fokker--Planck equation
\be
L^\dag \pi=0,
\ee
where
\be
L^\dag = -\nabla \cdot F+\frac{1}{2}\nabla\cdot D\nabla
\ee
is the Fokker--Planck generator with $D= \sigma\sigma^\mathsf{T}$ the diffusion matrix. The diffusion is not assumed to be reversible, which means that $\pi(x)$ is not necessarily a Gibbs density and that the stationary Fokker--Planck current
\be
J_{F,\pi}(x) = F(x)\pi(x) -\frac{D}{2}\nabla \pi(x)
\label{eqcurr1}
\ee
is not necessarily zero. The stationary current vanishes, as is well known, in the case where $n=m$, $\sigma$ is chosen to be proportional to the identity matrix, $\sigma=\epsilon \idf$, and the drift is conservative, $F=-\nabla U$. Then
\be
\pi(x) = c\, e^{-2U(x)/\epsilon^2},
\label{eqgibbs1}
\ee
where $c$ is a normalization constant. In this case, $X_t$ is a ``gradient'' diffusion that satisfies detailed balance with respect to $\pi$, leading to $J_{F,\pi}(x)=0$ for all $x$.

We are interested here in the fluctuations of time-integrated observables of $X_t$ and, more precisely, in the rate function of such observables characterizing their fluctuations in the long-time limit \cite{dembo1998,hollander2000,touchette2009,touchette2017}. We consider general ``current-type'' observables \cite{chetrite2014} of the form
\be
A_T = \frac{1}{T}\int_0^T g(X_t)\circ dX_t,
\label{eqobs1}
\ee
where $g$ is an arbitrary vector field on $\reals^n$ and $\circ$ denotes the Stratonovich (scalar) product corresponding to the mid-point integration rule \cite{jacobs2010}. The rate function is defined as 
\be
I(a) = \lim_{T\ra\infty} -\frac{1}{T}\ln P(A_T=a),
\ee
where $P(A_T=a)$ is the probability density of $A_T$, and is such that $I(a)\geq 0$ with equality at the stationary value 
\be
a_\pi = \int g(x)\cdot J_{F,\pi}(x)\, dx.
\ee
This implies that $P(A_T=a)$ is exponentially peaked around $a_\pi$; fluctuations around this concentration value are exponentially unlikely to occur, which means that $A_T$ converges to $a_\pi$ with probability 1 as $T\ra\infty$, in accordance with the ergodic theorem \cite{dembo1998,hollander2000,touchette2009}. 

The rate function of $A_T$ can be calculated using different methods, one of which involves solving for the principal eigenvalue of a linear differential operator related to $L^\dag$ \cite{touchette2017}. Another method consists in ``contracting'' the large deviations of the empirical density 
\be
\rho_T(x) =\frac{1}{T}\int_0^T \delta(X_t-x) dt
\ee
and the empirical current
\be
J_T(x) = \frac{1}{T}\int_0^T \delta(X_t-x)\circ dX_t
\ee
down to the observable $A_T$ using the fact that the joint large deviations of $\rho_T$ and $J_T$ have the explicit rate function
\be
I[\rho,j] = \frac{1}{2}\int (j-J_{F,\rho})\cdot (D\rho)^{-1} (j-J_{F,\rho})dx
\label{eq25rfct1}
\ee
if $\nabla \cdot j=0$. Since $A_T$ can be expressed as a function of $J_T$ as
\be
A_T =\tA(J_T)= \int g(x)\cdot J_T(x) dx,
\label{eqcont1}
\ee
we can then write, using the contraction principle \cite{touchette2009}, 
\be
I(a) = \inf_{\rho,j:\tA(j)=a} I[\rho,j],
\label{eqratecont1}
\ee
where the minimization is over all normalized densities and current fields such that $\tA(j)=a$ and $\nabla\cdot j=0$. Naturally, the global minimizers leading to the stationary value $A_T=a_\pi$ are $\rho=\pi$ and $j=J_{F,\pi}$, the stationary values of $\rho_T$ and $J_T$, such that $I[\pi,J_{F,\pi}]=0$.

This contraction of the so-called \emph{level 2.5 of large deviations} for Markov processes \cite{maes2008,maes2008a,barato2015,hoppenau2016} can be interpreted physically in terms of a modified process creating the fluctuation $A_T=a$, whose stationary density and stationary currents are the minimizers of (\ref{eqratecont1}), as explained in \cite{chetrite2015}. For our purpose, what is important to note is that the minimization problem (\ref{eqratecont1}) is very difficult to solve in general. In fact, it is as difficult to solve as the spectral problem reviewed in \cite{touchette2017}, since the spectral solution gives the solution to (\ref{eqratecont1}) and vice versa \cite{chetrite2015}. 
For this reason, it is natural to seek approximations of $I(a)$ by restricting the minimization (\ref{eqratecont1}) over a restricted class of solutions or by proposing sub-optimal solutions or ``ansatz.'' Because of the variational form of the problem, any such ansatz yields not just an approximation but an upper bound, since $I(a)\leq I[\rho,j]$ for any $\rho$ and $j$ satisfying the constraints. 

\section{Entropic bound}

The entropic bound is a general upper bound on the rate function $I(a)$ having the form
\be
I(a)\leq \frac{(a-a_\pi)^2}{4 a_\pi^2}\Sigma_\pi,
\label{eqbound1}
\ee
where $a_\pi$ is the stationary value of $A_T$ and $\Sigma_\pi$ is the stationary entropy production \cite{lebowitz1999}, given by
\be
\Sigma_\pi = 2 \int F(x)\cdot D^{-1}J_{F,\pi}(x)\, dx.
\ee
Since $A_T$, as defined in (\ref{eqcont1}), is a linear contraction of the empirical current, the bound is most often interpreted as applying to a scalar current (e.g, the space-integrated current), but other interpretations are possible depending on the form of $g$. For example, it can be interpreted as a bound on the  entropy production rate function by choosing $g=2D^{-1}F$, in which case $a_\pi = \Sigma_\pi$ \cite{pigolotti2017}. 

The bound can also be applied to reduced or coarse-grained currents (e.g., the empirical current along one coordinate), although $\Sigma_\pi$ is always the \emph{total} entropy production associated with all the coordinates or degrees of freedom of the system considered. This is important for experiments, which usually probe only a subset of the complete state-space. In general, it is known that reduced entropy productions are smaller or equal to the total entropy production \cite{celani2012,bo2014}, even if the degrees of freedom that are ``traced-out'' are associated with equilibrium dynamics. Therefore, the bound (\ref{eqbound1}) does not necessarily apply by replacing $\Sigma_\pi$ with the entropy production measured experimentally.

The entropic bound was conjectured in \cite{pietzonka2015} based on numerical results, proved in \cite{gingrich2016} for jump process, and later derived in \cite{gingrich2017} for diffusions by taking a diffusion limit of jump processes similar to the Kramers--Moyal expansion. Its direct proof from the contraction (\ref{eqratecont1}) is basically contained in \cite{polettini2016,gingrich2017} and is reproduced here for completeness. It simply follows by adopting the ansatz suggested in \cite{gingrich2017}, namely, 
\be
\hrho(x)=\pi(x),\qquad \hj (x) = \frac{a}{a_\pi} J_{F,\pi}(x).
\label{eqansc1}
\ee
The constraint $A_T=a$ is trivially satisfied by the global rescaling of the stationary current, since
\be
\tA(\hj) = \frac{a}{a_\pi} \int g(x)\cdot J_{F,\pi}(x)\, dx = a, 
\ee
and so is the divergence constraint, since $J_{F,\pi}$ is itself divergenceless. Inserting this ansatz into the level 2.5 rate function then yields the bound
\begin{eqnarray}
I(a) &\leq & I[\hrho,\hj] \nonumber\\
&=& \frac{(a-a_\pi)^2}{2 a_\pi^2} \int J_{F,\pi}\cdot (D\pi)^{-1} J_{F,\pi}dx\nonumber\\
&=& \frac{(a-a_\pi)^2}{2 a_\pi^2} \int \left(F-\frac{D}{2}\nabla\ln \pi\right)\cdot D^{-1}J_{F,\pi}\, dx\nonumber\\
&=& \frac{(a-a_\pi)^2}{4a_\pi^2} \Sigma_\pi, 
\label{mainres1}
\end{eqnarray}
the last equality following from integration by parts and the divergenceless current. As an extra result, it is easy to show that the bound satisfies the fluctuation relation \cite{gingrich2016} because $\hj$ is odd under time reversal and
\be
I[\rho,-j] = w[\rho,j] +I[\rho,j],
\ee
where $w$ is essentially the entropy production \cite{barato2015}.

The bound can also be understood by noting that, although the joint fluctuations of $\rho_T$ and $J_T$ are not Gaussian because of the term $J_{F,\rho}$ coupling $\rho$ and $j$ in the level 2.5 rate function, the current fluctuations for a \textit{fixed} density $\rho$ are Gaussian. Therefore, since any affine transformation of a Gaussian is also Gaussian, we must obtain a quadratic rate function for $A_T$ with its mean given by the mean current. 
This works here because $A_T$ is a contraction only of the current, so it is natural to keep $\rho$ as the stationary density $\pi$ and only scale the current as in (\ref{eqansc1}). This, however, is only a sub-optimal solution of the level 2.5 contraction: the real optimal solution changes both $\rho$ and $j$ from their stationary values in a way that is highly non-trivial, as discussed on general ground  in \cite{chetrite2015} and illustrated with examples in \cite{hoppenau2016}.

\section{Process interpretation}
\label{secproc}

We now come to our main result, which is to show that the ansatz (\ref{eqansc1}) associated with the quadratic entropic bound represents a physical diffusion with a modified drift. This can be seen in two ways. The first rests on the simple observation that, although many different Markov diffusions can have the same stationary density, there is a unique Markov diffusion, for a fixed $D$, that has a given stationary density and a given stationary current obeying (\ref{eqcurr1}). Therefore, the ansatz (\ref{eqansc1}) determines a unique Markov diffusion of drift $\hu$ and diffusion matrix $D$, with stationary density $\pi$ and stationary current 
\be
J_{\hu,\pi} = \hu \pi -\frac{D}{2}\nabla\pi=\hj,
\ee
leading to 
\be
\hu =  \frac{a}{a_\pi}F+\left(1-\frac{a}{a_\pi}\right) \frac{D}{2}\nabla \ln \pi.
\label{eqcontans1}
\ee
This can be re-expressed in a more suggestive way as
\be
\hu=\frac{a}{a_\pi} F_{\diss}+F_{\eq},
\label{eqcontans2}
\ee
where 
\be
F_{\eq}=\frac{D}{2}\nabla\ln \pi
\ee
is the reversible or \emph{equilibrium} component of the drift $F$ deriving, in a potential way, from the stationary density $\pi$ and
\be
F_{\diss} =F-F_{\eq} =\frac{J_{F,\pi}}{\pi}
\ee
is the non-reversible or \emph{dissipative} component of the drift, also called the current or hydrodynamic velocity, that controls the stationary current when $\pi$ is fixed \cite{nelson1967,chetrite2009,chetrite2011}. Thus we see that the ansatz (\ref{eqansc1}), at the process level, amounts to globally scaling the dissipative part of the total drift $F$ to produce the fluctuation $A_T=a$ while keeping the stationary density $\pi$.

The same result can be obtained more directly from a different variational representation of rate functions described in \cite{chetrite2015} (see also \cite{jack2015}), which considers the controlled diffusion
\be
dX_t^u = u(X_t^u)dt +\sigma dW_t,
\ee
and optimizes over all possible \emph{control drifts} $u$ leading to the fluctuation $A_T =a$ to obtain
\be
I(a) = \inf_{u:A_T=a} \frac{1}{2}\int (u-F) \cdot D^{-1}(u-F) \pi^u \,dx,
\label{eqcontvar1}
\ee
where $\pi^u$ is the stationary density of $X_t^u$. It can be verified that inserting the drift ansatz (\ref{eqcontans1}), which is such that $\pi^{\hu}=\pi$, into this variational formula yields the entropic bound (\ref{eqbound1}). The quadratic form of the bound follows here because the control problem (\ref{eqcontvar1}) becomes quadratic in $u$ when $\pi^u$ does not depend on $u$. 

The advantage of this approach is that we now have a direct interpretation of the entropic bound as being produced by a controlled process in which the dissipative component of the original drift is scaled by a constant. A control needs to be included in the original process to force it to reach the fluctuation $A_T=a$, and the entropic ansatz introduces this control onto the dissipative drift to keep $\pi$ as the stationary density. The functional on the right-hand side of (\ref{eqcontvar1}) is the control cost, corresponding mathematically to the logarithm of the Radon--Nikodym derivative of the controlled process with respect to the original process; see \cite{chetrite2015} for detail.

Another important advantage of seeing the entropic bound as being produced by a process is that we can understand more quantitatively why it is sub-optimal. The optimal drift solving the variational representation (\ref{eqcontvar1}) of $I(a)$ is known to be
\be
F_k = F +D(kg+\nabla \ln r_k),
\ee
where $g$ is the function defining the observable $A_T$, $r_k$ is the eigenfunction associated with the principal eigenvalue of a certain linear operator related to the large deviation problem (see \cite{touchette2017}), and $k$ is chosen according to $I'(a)=k$. 

This optimal drift is very difficult to find in general, just as the optimal density and current solving (\ref{eqratecont1}) are difficult to find \footnote{The optimal density of the contraction is known to be given by $\rho_k=l_k r_k$ where $l_k$ is the eigenfunction dual to $r_k$ \cite{chetrite2014}.}. Its explicit form clearly shows, however, that the optimal drift leading to $A_T=a$ is obtained by adding both a dissipative term to $F$ and a gradient, reversible term, corresponding to $D\nabla\ln r_k$, which changes the stationary density of the process. The entropic ansatz therefore make two approximations about the optimal drift, namely: it only changes the dissipative part of $F$, as mentioned before, and it is  a global scaling of the original dissipative force. None of these assumptions is true, in general, and so must necessarily lead to an upper bound on the true rate function $I(a)$. 

This can be illustrated simply for the diffusion on the unit circle defined by
\be
d\theta_t = (\gamma -U'(\theta))dt+\sigma dW_t,
\label{eqring1}
\ee
where $\theta_t\in [0,2\pi)$, $U(\theta)$ is a periodic potential, $\sigma>0$ is the noise intensity, and $\gamma>0$ is a constant that drives the diffusion in a nonequilibrium steady state \cite{reimann2002}. The optimal control process, also called the \emph{driven} or \emph{effective process}, was constructed recently in \cite{tsobgni2016} for the space-integrated current
\be
J_T =\frac{1}{T}\int_0^T d\theta_t,
\ee
which corresponds to the choice $g=1$. The results show that small current fluctuations close to the stationary current are optimally created by modifying not just the drive, but also the potential $U(\theta)$ in a nonlinear way (see Fig.~5 of \cite{tsobgni2016}). Consequently, the stationary density is modified, which shows that any ansatz that does not change the stationary density is sub-optimal in general compared to the driven process. 

This applies for fluctuations of $A_T$ arbitrary close to the stationary value $a_\pi$, so we also conclude that the drift of the entropic bound does not correspond in general to a linear perturbation of the optimal drift around $F$. In other words, although the drift $\hu$ is linear in $a$, it does not correspond in general to a linear perturbation of $F_k$ around $k=0$, which gives the Taylor expansion of $I(a)$ to second order in $a$ and therefore the variance of $A_T$ \footnote{In practice, this Taylor expansion can be constructed as a linear perturbation of the spectral problem defining $r_k$.}. This explains why the entropic bound gives a lower bound to the true asymptotic variance of $A_T$.

The only case where the entropic ansatz is optimal for the ring model is when $\gamma$ or $\sigma$ becomes very large compared to the potential height. Then the optimal driven process has a near constant density, which is the stationary density of the diffusion (\ref{eqring1}) without potential, and only scales the drive $\gamma$ so as to reach different current fluctuations. As this reproduces the entropic ansatz, we find a quadratic rate function for the current. The same result was obtained in a much more complicated way recently by studying the housekeeping heat \cite{hyeon2017}.

To close, we note that the entropic process with drift $\hu$ has a stationary entropy production equal to
\be
\hsig = 2\int \hu\cdot D^{-1}\hj\, dx = \frac{a^2}{a_\pi^2}\Sigma_\pi.
\ee
Moreover, it can be verified that the entropic process for $A_T=-a_\pi$ corresponds to the time-reversal of the original diffusion (\ref{eqsde1}) \cite{haussmann1986}, which has the same stationary density $\pi$ but whose drift is 
\be
F'= -F+D\nabla\ln\pi.
\ee 
It is known that this is also the optimal driven process related to $A_T=-a_\pi$, by reversal of the current \cite{bonanca2016}. Therefore, the entropic process is optimal for $A_T=-a_\pi$, which explains physically why the bound is tight at that point and why the bound satisfies overall the fluctuation relation. Of course, it is also optimal for $A_T=a_\pi$ because $\hu = F_{k=0} = F$.

\section{Reversible limit}

The entropic drift $\hu$ does not generally correspond, as we have seen in the previous section, to a linear perturbation of the optimal drift $F_k$ for small fluctuations of $A_T$ around its stationary value $a_\pi$. Another property of $\hu$ and the entropic bound is that they can be, in general, non-perturbative: that is, if we consider a perturbation $F+\lambda G$ of the original diffusion with drift $F$, then $\hu$ and the bound may involve $G$ rather than $\lambda G$.

To illustrate this, we consider a simple perturbation of a gradient diffusion with $\sigma=\epsilon\idf$ and drift given by
\be
F(x) = -\nabla U(x) +\lambda G(x),
\ee
where $\lambda$ is a small parameter multiplying a non-gradient perturbation $G$, chosen such that $G\cdot \nabla U=0$ and $\nabla\cdot G=0$. Under these ``transverse'' conditions, the stationary density $\pi$ is still the Gibbs density (\ref{eqgibbs1}) \cite{risken1996}, although there is now a non-zero current given by $J_{F,\pi} = \lambda G \pi$ to first order in $\lambda$, so that $a_\pi=\lambda \langle g\cdot G\rangle_{\pi}$, where $\langle\cdot\rangle_\pi$ denotes the expectation over $\pi$. Moreover, it can be checked that the stationary entropy production is
\be
\Sigma_\pi= 2\lambda^2 \langle G\cdot D^{-1} G\rangle_{\pi}
\ee
to lowest order in $\lambda$. As a result, we obtain for the entropic bound
\be
I(a)\leq \frac{a^2}{2} \frac{\langle G\cdot D^{-1}G\rangle_{\pi}}{\langle g\cdot G\rangle^2_{\pi}}+O(\lambda).
\label{eqbrev1}
\ee

The same result can be obtained more directly by noting that the entropic drift is
\be
\hu = \frac{a}{a_\pi} \lambda G-\nabla U = \frac{a}{\langle g\cdot G\rangle_\pi} G-\nabla U
\ee
at lowest order in $\lambda$. As both results are non-perturbative in $G$, the entropic bound is not expected to be a meaningful bound for reversible systems, unless the expectations on the right-hand side of (\ref{eqbrev1}) involving $G$ and $g$ cancel. Indeed, if $I(a)$ exists for these systems, then it should depend only on the reversible (i.e., gradient) part of the drift and have corrections at order $\lambda^2$ coming from the non-reversible perturbation, which is not seen for the entropic bound. It could be, of course, that $A_T$ does not fluctuate in the reversible limit, in which case $I(a)$ and its bound become degenerate. 

This happens, for example, when considering the entropy production of gradient systems, which depends only on their final and initial conditions in a way that is not covered by the Level-2.5 large deviations. The ring model provides, on the other hand, an example for which the rate function $I(a)$ is well defined in the reversible limit where $\gamma\ra 0$ and is bounded by an entropic bound that does not depend on the driving $\gamma$, since $G=\gamma$ and $g=1$, so that the two expectations in (\ref{eqbrev1}) do cancel. It is clear that we should not expect this cancellation in more general systems involving space-dependent perturbations. If we add a potential to the ring model, for example, then the current function is not in general quadratic \cite{tsobgni2016}, and so it is not saturated by the entropic bound even at equilibrium ($\gamma=0$) \cite{polettini2016}.

\section{Concluding remarks}

We have shown that the entropic bound recently derived for current rate functions can be interpreted in terms of an effective process that makes the fluctuation typical. The entropic bound is easy to construct in practice, as it involves only stationary quantities (current and entropy production) that can be measured experimentally from steady-state averages. It is also the simplest bound that can be found by observing that the current fluctuations are Gaussian when the density is fixed or, from the process perspective, when the non-reversible part of the drift is scaled independently of the reversible part. 

We expect this process interpretation of the entropic bound to be useful in the future to construct or approximate the driven process that creates a given fluctuation $A_T=a$ at the optimal cost given by rate function $I(a)$. The drift of this process is known explicitly, but is difficult to obtain, as mentioned. One way to approach it could be to start with the simple form of the entropic drift and consider perturbations of the reversible drift that lower the control cost given in (\ref{eqcontvar1}) compared to the quadratic bound. This is essentially a control problem, which can be solved using many standard techniques; see \cite{chetrite2015} for details. In principle, the driven process can also be constructed for general discrete- and continuous-time Markov chains \cite{chetrite2014}. However, it is not clear for these that the  entropic bound can be interpreted as being realized by an effective (sub-optimal) Markov chain, since there is an additional approximation of the level 2.5 rate function involved in the derivation of the bound compared with diffusions \cite{gingrich2017}. This is an open problem.

Another interesting problem is to derive quadratic bounds for more general observables having the form
\be
A_T = \frac{1}{T}\int_0^T f(X_t)dt+\frac{1}{T}\int_0^T g(X_t)\circ dX_t,
\label{eqobs2}
\ee
which involves a time integral depending on a function $f$ of the process in addition to the current part considered before. The contraction (\ref{eqratecont1})  in this case involves not only the empirical current, but also the empirical density via the constraint
\be
A_T=\tA(\rho_T,J_T) = \int f(x)\, \rho_T(x)\, dx+\int g(x)\cdot J_T(x)\, dx,
\ee
which means that it can now be approximated by changing either the density or the current or both, leading to different upper bounds on the rate function $I(a)$. 

Whether such bounds are quadratic is unknown. What is clear is that there cannot be a \textit{universal} quadratic upper bound like the entropic bound that applies to all observables $A_T$ having the form (\ref{eqobs2}). For instance, the rate function of the empirical variance of the Ornstein--Uhlenbeck process, corresponding to the choice $f(x)=x^2$ and $g=0$, is known to diverge like $1/a$ as $a\ra 0$ \cite{bryc1997}, and so cannot be bounded by a parabola. However, there can be sub-optimal processes similar to the entropic process that give quadratic upper bounds for specific observables or in certain ranges of fluctuations. Such processes would give, if derived around the stationary value $A_T=a_\pi$, new uncertainty relations constraining the variance of $A_T$.

\begin{acknowledgments}
We thank Todd Gingrich and Grant Rotskoff for comments on this work. We are also grateful for the hospitality of ICTS-TIFR Bangalore, where this work was done. HT is supported by the  National Research Foundation of South Africa (Grant no 96199). CN is supported by an Aide Investissements d'Avenir du LabEx PALM (ANR-10-LABX-0039-PALM).
\end{acknowledgments}

\bibliography{masterbib}

\end{document}